\documentclass[12 pt]{article}
\usepackage{graphicx} % Required for inserting images

\usepackage{sectsty}
\usepackage[italicdiff]{physics}
\usepackage[margin=1in]{geometry}
\sectionfont{\normalsize \centering}
\subsectionfont{\normalsize	\centering}

\usepackage{amsmath}

\usepackage{setspace}
\usepackage{natbib}
\AtBeginDocument{} % no &, please!

\makeatletter
\newcommand{\gobblecomma}{\@ifnextchar,{\@gobble}{}}
\makeatother

\AtBeginDocument{}

\renewcommand{\cite}{\citer}
\newcommand{\citer}[1]{\citeauthor{#1} ([\citeyear{#1}])}

\renewcommand{\citet}{\citeparen}
\newcommand{\citeparen}[1]{\citeauthor{#1} [\citeyear{#1}]}

\usepackage{xcolor}
\usepackage{soul}

\sethlcolor{yellow}

\renewenvironment{abstract}{
  \par\noindent\rule{\textwidth}{0.1pt}\par\vspace{0.2em}% thin top line, minimal space
  \begin{spacing}{1.15}
  \noindent\fontsize{11}{13}\selectfont  % 11 pt text, ~13 pt baseline skip
}{
  \end{spacing}
  \par\vspace{-0.2em}\noindent\rule{\textwidth}{0.1pt}\par % thin bottom line
}

\usepackage{footmisc}

\usepackage{amsthm}

\title{Limiting Reduction and Modified Gravity}
\author{\Large
  Antonis Antoniou\footnote{Department of History and Philosophy of Science, National and Kapodistrian University of Athens, Athens, Greece, contact: antantoniou@phs.uoa.gr}
  \and
  \Large Lorenzo Lorenzetti\footnote{Institute of Philosophy, Università della Svizzera italiana (USI), Lugano, Switzerland, contact: lorenzo.lorenzetti.ac@gmail.com}
}
\date{\normalsize Forthcoming in \textit{The British Journal for the Philosophy of Science}}

\begin{document}

\maketitle

\begin{abstract}

\noindent Modified Newtonian Dynamics (MOND) is a framework of theories that adjust Newton's laws of gravity to explain effects such as galactic rotation anomalies, offering an alternative to dark matter. This essay examines the justification of MOND by assessing its inter-theoretical relationship to established theories across relevant scales, in particular its connection to Newtonian gravitation. We argue that MOND fails a key condition for a theory's justification---what we call `reduction-wise justification'---since it does not adequately reduce to Newtonian gravity in a fully non-arbitrary way. More precisely, despite satisfying the standard formal criteria for successful limiting reduction, MOND does not properly reduce to Newtonian gravitation because of (i) the absence of a fundamental theoretical framework to justify the interpolating function introduced in MOND and (ii) the lack of a unified mathematical structure working across all scales, independent of Newtonian theory. Hence, the case study of MOND provides crucial results for the general debate on inter-theoretic reduction in science: MOND’s failure as a case of reduction highlights important limitations in standard accounts of limiting reduction. We respond by proposing a more refined framework for limiting reduction that introduces two additional criteria to better distinguish successful from pathological reductions. More broadly, this case illustrates how analysing reduction-wise justification can serve as a powerful tool for evaluating the validity of novel theories that are not yet empirically established.

\end{abstract}

\section{Introduction} 

Modified Newtonian Dynamics (MOND) is a framework for theories of gravity where gravitational dynamics deviate from Newtonian dynamics within a certain ultra-low-acceleration regime, and is often presented as a possible alternative to the postulation of non-baryonic dark matter. It was first proposed by Milgrom in 1983 as a possible explanation of the unexpected flat rotation curves of stars in the outskirts of galaxies, potentially eliminating the need for dark matter. Since then, several attempts have been made to implement Milgrom's initial hypothesis into more complete classical and relativistic theories of gravity, while at the same time recovering the effective success of Newtonian dynamics and general relativity as limiting cases. Today, despite the remarkable success of MOND theories at the galactic scale, they still face a number of important challenges and are widely considered, even by their proponents, as incomplete.

This paper has a threefold aim. The first, more direct, is to assess the validity of theories within the MOND framework by determining whether they are properly related to the theories they are supposed to be reducible to. The answer is negative: they are not fully `\textit{reduction-wise justified}', failing to satisfy a crucial condition for a theory’s justification. We conclude that MOND is not (yet) justified as a framework. MOND also serves as an excellent case study for expanding our understanding of how reduction works and its advantages. Indeed, building on our analysis of MOND, the second goal is to show that assessing the reduction-wise justification of a theory is a powerful tool for analysing whether a new, contested theory—one not yet established—is justified. The third and most ambitious goal is to use MOND’s reduction as a case study to develop a more refined version of the limiting approach to inter-theory reduction. Limiting reduction is a key account of theoretical reduction, used by scientists and philosophers to model scientifically informed cases of inter-theory reduction. Our analysis shows that standard limiting reduction is not fine-grained enough to distinguish valid from invalid cases. Hence, we propose a novel version that improves on the standard one.

The article is structured as follows. Section 2 explains why theoretical reduction is crucial to the debate around a theory's justification and introduces limiting reduction. Section 3 introduces MOND and delves into the details of the relationship between MOND and the topic of inter-theoretic reduction. Section 4 asks whether MOND is a valid case of limiting reduction and can be regarded as a reduction-wise justified theory. To respond we apply the three formal criteria set by standard limiting reduction: (a.) \textit{limit}: a theory's quantities reduce to corresponding quantities of a narrower theory by taking an appropriate upper or lower limit of a parameter; (b.) \textit{physicality}: the limiting operation is physically realistic and involves only regular limits; (c.) \textit{non-arbitrariness}: the limiting parameter is not arbitrarily introduced and can be given a meaningful physical interpretation. We show in detail that MOND satisfies these criteria, although the third deserves more attention. More precisely, we argue that the limit used to reduce MOND to Newtonian gravity in the appropriate regime is taken on a meaningful quantity and is therefore not arbitrary in an important sense. However, it is arbitrary in that the exact form of the function on which the limit is applied is introduced specifically to recover Newtonian gravity in the limit. This would render MOND reduction-wise unjustified—if not for the fact that, at least prima facie, other canonical and uncontroversial instances of reduction seem to operate limits in a strikingly similar arbitrary manner. This highlights a crucial observation: the criterion of non-arbitrariness does not appear fine-grained enough to pinpoint the precise reasons for the invalidity of a given case of reduction.

Section 5 builds on this consideration and achieves two results. First, it compares the reduction of MOND laws to Newtonian gravity with the reduction of Planck's law of radiation to Wien’s and Rayleigh’s laws. Like MOND, Planck’s law was introduced by interpolation and reduces to Wien’s and Rayleigh’s laws in a similarly arbitrary manner. However, despite these similarities, we show that there is a crucial reason why MOND’s case is invalid, while Planck’s is valid. Specifically, although the form of Planck’s law was arbitrarily introduced, it was later supported by deeper theoretical grounds—most notably, the development of quantum theory, which justifies its exact formulation. This more fundamental backing is (not yet) present in the case of MOND. Hence, second, to clarify this distinction more rigorously, we refine the non-arbitrariness criterion by implementing a further condition that allows us to successfully differentiate the two cases analysed in this section: \textit{`downward theoretical grounding'}.

Section 6 expands our analysis of the unjustifiedness of MOND reductions and our refinement of the general framework of limiting reduction by introducing a further comparison: the reduction of general relativity to Newtonian gravity. The (newly introduced) key criterion that general relativity satisfies—but MOND does not—is yet another dimension in which condition of non-arbitrariness can be specified, which we call \textit{`upward theoretical grounding'}. This requires that the mathematical framework of the reducing (narrower) theory be naturally embedded within the mathematical framework of the reduced theory (the broader theory) and that the latter should, in principle, account for phenomena at all scales without having to rely on the framework of the narrower theory.

Section 7 concludes by presenting and justifying, in a unified way, the refined framework for limiting reduction that emerges from our detailed discussion of MOND. This framework incorporates the two original criteria (a.)–(b.) along with a newly refined version of condition (c.), comprising the two new conditions, which further specify the non-arbitrariness criterion. We propose an understanding of criterion (a.) as a necessary and defining condition for limiting reduction, while criteria (b.)--(c.) are further conditions that can strengthen the case for successful limiting reduction. These refinements clarify what renders a theory such as MOND reduction-wise unjustified, and at the same time explain why, by contrast, general relativity and Planck’s law are more uncontroversially regarded as justified cases of limiting reduction.

\section{Limiting Reduction}

\subsection{Why reduction matters}

Establishing inter-theory reduction between theories is a central part of scientific practice and has significant implications (Crowther [\citeyear{Crowther2020-CROWIT-3}], Palacios [\citeyear{Palacios2023-PALIRI-2, sep-physics-interrelate}]). Broadly put, reduction is a relation between two theories (or parts thereof) wherein a narrower, less accurate, or less fundamental theory can be derived from a broader, more accurate, or more fundamental one. Here we focus in particular on reduction within physics.

Given our goal of assessing MOND's validity, there is one particular aspect of inter-theoretic reduction we are interested in: the connection between reduction and the justification of a theory. In particular, we consider inter-theoretic reduction a necessary (but not sufficient) condition for the justification of a theory when it is introduced in a domain where a more narrowly scoped theory is already well-established.\footnote{On the other hand, we acknowledge that a new theory could be formulated in a domain where no prior theory exists, in which case its justification must be sought by other means.} That is, establishing inter-theoretic reduction is a necessary condition for the justification of a newly introduced and more fundamental theory by showing that an already established, less fundamental theory of the same domain is appropriately recovered from the more fundamental one at the expected scale. For instance, the reducibility of general relativity to Newtonian gravity is a necessary condition for the justification of general relativity as a broader theory of gravity, by virtue of showing how and why the theory applies to a broader domain compared to Newtonian gravity and that the latter is indeed an effective theory in the limited domain of non-relativistic velocities. By the same token, a theory of quantum gravity can only be considered as a viable candidate insofar as it recovers the theory of general relativity in the appropriate non-quantum regime. Thus, demonstrating the correct reductive relationships is a necessary condition for the establishment of new, more fundamental theories and we shall say that every theory satisfying this condition is reduction-wise justified. 
One of our main concerns in this article is the extend to which classical and relativistic theories of MOND are indeed justified in terms of this particular aspect of inter-theoretic reduction.\footnote{Terminology varies: the more fundamental theory is sometimes called reducing theory and the less fundamental reduced theory, though labels may be inverted. We follow standard limiting reduction convention, calling the broader theory (MOND) the reduced theory (Nickles [\citeyear{nickles1973}]).}

In what follows, we focus on the limiting account of reduction, which fits well within our context and sets clear standards for assessing whether an instance of reduction qualifies as successful. As noted earlier, if the reductive relationships of a theory are problematic, its justification is called into question. This section introduces a standard account of limiting reduction and the conditions it imposes on successful instances of reduction by specifying certain criteria under which a given limiting operation supports a claim of reduction. The approach is applied here to the standard example of the reduction of relativistic momentum to classical momentum.

\subsection{Introducing limiting reduction}

One standard account of theoretical reduction in physics is Nagelian reduction (Nagel [\citeyear{Nagel1961}], Palacios [\citeyear{sep-physics-interrelate}]). In this framework, reducing a theory \( T_2 \) to a theory \( T_1 \) requires deriving the laws of \( T_2 \) from those of \( T_1 \), assisted by bridge laws that link the terms of \( T_1 \) and \( T_2 \) when the theories use different vocabularies. Another major account, introduced by \cite{nickles1973}, similarly characterizes theory reduction as law-derivation but frames it in terms of mathematical operations such as limits and approximations. Specifically, this approach emphasizes inter-theory operations performed on particular quantities, enabling the derivation of the quantities of the reduced theory from those of the reducing theory, rather than focusing on the full deduction of \( T_2 \)’s laws from those of \( T_1 \). A special case of Nickles' reduction which is particularly well-suited for our purposes is \textit{limiting reduction}, where the applied mathematical operations are limiting operations (Palacios [\citeyear{palacios2022}], p. 56):

\begin{quote}
    \textbf{Limiting reduction}: Let $Q^1$ denote a relevant quantity of $T_1$, $Q^2$ a relevant quantity of $T_2$, then a quantity $Q^2$ of $T_2$ \textit{reduces}$_{lim}$ to a corresponding quantity $Q^1$ of $T_1$ iff (i) lim$_{N\rightarrow \infty}$$Q^1_N = Q^2$ or lim$_{N\rightarrow 0}$$Q^1_N = Q^2$ (where $N$ represents a parameter appearing in $T_1$) and (ii) the limiting operation makes physical sense.
\end{quote}

This characterisation of theoretical reduction naturally accounts for several standard instances of reduction. Consider a simple example to evaluate conditions (i) and (ii), which will be useful later for discussing limiting reduction in MOND.

Let's consider condition (i) first. We shall name this condition `\textit{limit}'. A paradigmatic example is the reduction of the relativistic expression for the momentum to the corresponding formula for momentum in classical mechanics. That is, the relativistic expression \( p = (m_0 v) / \sqrt{1 - (v^2 / c^2)} \) converges to the classical expression \(p=m_0v\) when we take the limit \((v/c)^2\rightarrow0\). Hence this operation satisfies \textit{limit}.

To examine whether it also satisfies condition (ii), that is, whether it makes physical sense, one may ask under which conditions the mathematical operation of taking a particular limit can be justified. Nickles provides the following explanation:

\begin{quote}
    It is important that the reductive operations make physical sense. By taking the limit as temperature T goes to infinity, we may be able to eliminate T or a more complex factor involving T from some theory. But an infinite temperature does not make physical sense. Nor does it make physical sense to allow T to vary normally but to eliminate it by introducing a special new multiplicative factor to which no physical interpretation can be assigned and on which the limits are taken. (Nickles [\citeyear{nickles1973}], pp. 200-201)
\end{quote}
We can thus identify the following two conditions. First, \textit{physicality}: an instance of reduction is justified insofar as the limiting operation does not involve the assumption of infinite values for quantities, as quantities can only take finite values in the actual world. Analogously, we might say that it can only be warranted if the limit does not take a parameter to zero, in cases where we know that in the physical world the value of the parameter is small but non-zero. Second, \textit{non-arbitrariness}: an instance of reduction of \(T_1\) is justified insofar as it does not involve the introduction of new arbitrary factors to which we cannot give a physical interpretation and to which limits are taken to recover \(T_2\).

To address Nickle's concerns about the introduction of infinities in limiting reduction, \cite{butterfield2011} has put forward a criterion for the justification of limits that has since been picked up by various authors (for instance, Landsman [\citeyear{Landsman2013-LANSSB-3}]). \cite{palacios2022} takes this principle as a way to specify the second criterion of Nickles' limiting reduction formally. The idea is that the result of taking a limit of a parameter represents realistic behaviour if it approximates the result for large but finite values of such a parameter (or for small but nonzero values). In that case, even an infinite limit---a mathematical artefact---would make physical sense, as it approximates a realistic result. Hence the criterion provides a way to formally specify the criterion of \textit{physicality} by deflating the concern about infinite limits. Following \cite{palacios2022}, the principle can be precisely expressed as follows (Butterfield [\citeyear{butterfield2011}], §3.3.3):

\begin{quote}
\textbf{Butterfield's Principle:} A limit is justified as being mathematically convenient and empirically adequate if the values of the quantities evaluated in the limit at least approximate the values of the quantities “on the way to the limit,” that is, for large but finite values \(N_0\) of the parameter \(N\), that is, if \(Q_\infty \approx Q_{N_0}\) and if the behavior “on the way to the limit” is the one that is physically real.
\end{quote}

Note that the principle can be equally applied to \(N \rightarrow 0\) limits, by requiring that the values of the quantities evaluated at the zero limit at least approximate the values of the quantities for small but non-zero values \(N_0\) of the limiting parameter \(N\). A limit that satisfies Butterfield's principle therefore automatically satisfies the condition of physicality.

Butterfield's principle can also be formulated in terms of regular and singular limits. Roughly speaking, a regular limit is a limit in which the nature of the solution near the limit is similar and close to the solution at the limit, that is a limit in which $Q_{\infty} \approx Q_{N_0}$ for very large but finite $N_0$ or $Q_0 \approx Q_{N_0}$ for very small $N_0$. Regular limits therefore satisfy Butterfield's principle by definition.\footnote{Batterman ([\citeyear{batterman2001}], p. 79) defines regular limits in a similar way.} A singular limit, on the contrary, is one in which the nature of the solution near the limit is fundamentally or structurally different from the nature of the solution at the limit, and therefore such limits are often considered unphysical. Mathematically, a clear way to show that a limit is regular is by showing that the function in consideration can be Taylor expanded in such a way that each successive term smoothly approaches zero, ensuring that the zeroth-order term corresponds to the limiting behavior, while higher-order terms represent progressively smaller corrections.\footnote{On the connection between Taylor expansion and regular limits see Bokulich ([\citeyear{Bokulich2008-BOKRTQ}], p. 16).} Once again, the provided example satisfies the condition since momentum in special relativity can indeed be reformulated in terms of the classical momentum plus an expansion in powers of \((v/c)^2\) since \(\sqrt{1-\frac{v^2}{c^2}}\) can be expanded in a Taylor series: 

\[1-1/2(v/c)^2-1/8(v/c)^4-1/16(v/c)^6-\dots\] 

Regular limits are uncontroversially regarded in the literature as suitable for reduction, as opposed to the disputed cases of singular limits, where lim$_{N\rightarrow \infty}$$Q^1_N \neq Q^2_{N=\infty}$ (see Berry [\citeyear{berry1995asymptotics}], Batterman [\citeyear{batterman2001}], Bokulich [\citeyear{Bokulich2008-BOKRTQ}]). For instance, a canonical case of a singular limit is often discussed in the context of first-order phase transitions (Batterman [\citeyear{batterman2001, Batterman2004-BATCPA}]). First-order phase transitions in thermodynamics are characterised by non-analytic (discontinuous) behaviour in the derivative of the free energy function, and the thermodynamic limit of the number of particles and volume \(N, V \rightarrow \infty\) is needed to derive the discontinuity. However, Batterman argues that this limit is singular because for an arbitrarily large but finite $N$, the derivatives of the free energy function are not discontinuous. In his own words: ``The behavior at the limit (the physical discontinuity, the phase transition) is qualitatively different from the behavior as that limit is approached." (Batterman [\citeyear{Batterman2004-BATCPA}], p. 236). Because of the presence of singular limits, he then claims that first-order phase transitions are in fact failing cases of limiting reduction.\footnote{However, note that several authors have disputed this conclusion by arguing that first-order phase transitions do not involve singular limits (Butterfield [\citeyear{butterfield2011}], Palacios [\citeyear{Palacios2019-PALPTA-4}]).}

As for the condition of \textit{non-arbitrariness}, this can be easily met by requiring that the introduced limiting operation does not involve the introduction of any uninterpreted and arbitrary parameter on which we take the limit to recover \(T_2\) from \(T_1\). The reason is quite straightforward. The introduction of such arbitrary limiting factors would make every reduction a trivial exercise, obscuring the physical interpretation of the reduced theory, especially with respect to the limiting parameter. The given example of the reduction of momentum clearly satisfies this condition since no arbitrary or uninterpreted factor is introduced to recover the classical expression \(p=m_0v\) from the relativistic one via the limit. Rather, every quantity involved has a perfectly meaningful interpretation and the taken limit resembles a situation in which the velocity condition is much smaller compared to the speed of light. The reduction of relativistic to Newtonian momentum also clearly satisfies conditions (i) and (ii) as specified above, and is therefore a clear case of a successful and justified limiting reduction.

In sum, the presented analysis in this section outlines what we consider the standard account of limiting reduction as a necessary---albeit not sufficient---condition for the justification of a theory. Namely, we have shown that a physical theory is reduction-wise justified in terms of mathematical limits insofar as the following conditions are satisfied: (a.) \textit{Limit}: its quantities reduce to corresponding quantities of a narrower theory by taking an appropriate upper or lower limit of a parameter; (b.) \textit{Physicality}: the limiting operation is physically realistic and involves only regular limits; and (c.) \textit{Non-arbitrariness}: the limiting parameter is not arbitrarily introduced and can be given a meaningful physical interpretation. We now proceed to present the framework for MOND and argue that MOND theories satisfy these criteria and can therefore be considered as justified theories in terms of reduction according to the standard account, even though they face important challenges and are widely considered as incomplete.

\section{Modified Newtonian Dynamics} \label{MOND}

\subsection{Outline}

The theoretical framework of Modified Newtonian Dynamics was first introduced by \cite{milgrom1983a} as an alternative hypothesis to dark matter. The core idea of Milgrom's proposed toy model for MOND was that standard Newtonian gravity takes a different form in the ultra-weak-field regime of low accelerations below a critical acceleration constant $a_0$, which was--rather unexpectedly--found in various unrelated scaling relations such as the Baryonic Tully-Fisher Relation (BTFR).\footnote{The BTFR is an empirical relationship between the baryonic mass of a galaxy and its asymptotic rotational speed. For more on these relationships and their connection to $a_0$ see Milgrom ([\citeyear{milgrom2020mond}], Sec. 3). For further historical/philosophical analyses of the debate between dark matter and MOND see \cite{sanders2010}, \cite{deswart2017}, and \cite{antoniou2025}.} The aspiration was (and still is) that this new acceleration constant would play a similar role as the Planck constant, $h$, in quantum mechanics and the speed of light, $c$, in special relativity by virtue of being the physical quantity whose limit determines the different form of gravitational dynamics between the weak-field regime (that is, the Newtonian limit) and the ultra-weak-field regime (that is, the deep-MOND limit). For large accelerations, $g \gg a_0$, Newtonian dynamics would be preserved, while for very low accelerations below the acceleration constant, $g \ll a_0$, the gravitational acceleration would take the MOND-ian form:

\begin{equation}\label{MONDaccel}
    g_M = \sqrt{g_{N}a_0}
\end{equation}

In more formal terms, the idea is that taking the limit $a_0 \rightarrow 0$ in the proposed theory should reproduce standard gravitational dynamics just like taking the limit of $h \rightarrow 0$ in quantum mechanics recovers classical mechanics. Similarly, taking the limit $a_0 \rightarrow \infty$ leads to the ultra-weak-field regime where the proposed modified gravitational law describing the observed galactic dynamics applies.\footnote{Since $a_0$ is a constant, saying that one takes the limit of this quantity, as well as the limit of other constants like $h$ and $c$, is somewhat misleading. What happens is that one takes the limit of a dimensionless quantity formed by the ratio of such constants to some other quantity of the same dimensions. In MOND theories, this ratio is denoted by the limiting parameter $x = g/a_0$.} To ensure the smooth transition between the weak-field (Newtonian) regime and the ultra-weak-field (MOND-ian) regime Milgrom formulated his phenomenological law of modified gravity, which he described as an `effective working formula' (Milgrom [\citeyear{milgrom1983a}], p. 366), in terms of an interpolating function $\mu(x) $ as follows:

\begin{equation}\label{Milgrom'sLaw}
   \mu \left( \frac{g}{a_0} \right) g = g_N 
\end{equation}
with the interpolating function satisfying the following mathematical criteria: (1) $\mu(x) \rightarrow 1$ for $x \gg 1$; (2) $\mu(x) \rightarrow x$ for $x \ll 1$. The first criterion ensures that in the weak-field regime, where $g \gg a_0$, gravitational acceleration remains the same as the Newtonian constant. The second ensures that in the ultra-weak-field regime, where $g \ll a_0$, the formula yields the desired modified gravitational acceleration to fit the relevant observations in galactic dynamics.

These constraints allow for significant flexibility in the exact shape of the interpolating function in theories of MOND. As a result, several different families of functions which can equally fit---to a greater or lesser degree---the available data from the weak gravity regime of galaxies and the strong gravity regime of the solar system have been proposed since Migrom's initial formulation of MOND. Up to this day, the cornerstone of every attempt to implement Milgrom's hypothesis into a theory of modified gravity is to reproduce Milgrom's effective working formula (\ref{Milgrom'sLaw}) in highly symmetrical systems with an interpolating function asymptotically obeying the aforementioned conditions. As will be shown, the role of the interpolating function is to achieve a smooth transition between a classical MOND theory for very low accelerations and the Newtonian limit, as well as between the relativistic versions of MOND and the weak and ultra-weak limit of the Newtonian and MONDian regime, respectively. 

The first serious attempt to implement Milgrom's effective law in a classical theory of modified gravity is found in the Aquadratic Lagrangian theory (AQUAL) by \cite{bekenstein1984}. The core idea in AQUAL is to alter the gravitational action and obtain a modified Poisson equation for a set of particles with mass $m_i$.\footnote{In general, the Poisson equation is a partial differential equation that relates a scalar field to the distribution of its source. In Newtonian gravity the equation takes the form $\nabla^2\Phi = 4\pi G \rho$ where $\Phi$ is the gravitational potential, $G$ is the gravitational constant and $\rho$ is the mass density.}  Varying the new action with respect to the gravitational potential $\Phi$ yields a modified Poisson equation:

\begin{equation}\label{ModPoisson}
    \nabla.\left[ \mu \left(\frac{|\nabla\Phi|}{a_0}\right) \nabla\Phi \right] = 4\pi G \rho
\end{equation}
where $\mu(x) = F'(z)$ is the required interpolating function, with $z=x^2$ obeying a new set of similar constraints to \ref{Milgrom'sLaw}. Solving (\ref{ModPoisson}) yields the new MONDian relation between the MONDian acceleration ($g_M=-\nabla\Phi$), and the Newtonian acceleration ($g_N=-\nabla\Phi_N$) which is equivalent to Milgrom's effective law.

It is widely acknowledged within the MOND community that such classical theories of modified gravity are merely toy theories for galactic-scale predictions, and that a more fundamental theory (FUNDAMOND) is required, within which the classical theory will appear as a limit. At the same time, classical theories provide useful weak-field targets for relativistic MOND theories. However, since MOND is acceleration-based, constructing a relativistic extension is challenging: in general relativity, the Christoffel symbols that play the role of acceleration are not tensors, so covariant formulations cannot rely on them. The solution is to replace acceleration with tensorial quantities, typically requiring additional fields as in scalar–tensor theories, leading to multifield relativistic models where the weak-field limit takes the generalised form: $ \Phi = \Phi_N + \phi $, where $\Phi_N$ obeys the Newtonian Poisson equation and the scalar field $\phi$ acts as a `phantom dark matter' potential governed by a modified Poisson equation.

A clear example of such an attempt comes from the relativistic extension of Bekenstein and Milgorm's theory, RAQUAL (Bekenstein and Milgrom [\citeyear{bekenstein1984}]), where the added scalar field plays the role of an auxiliary potential whose gradient has the dimensions of acceleration. To recover the required MOND dynamics, the standard Einstein-Hilbert action is used along with a \textit{k-essence} action for the added scalar field inspired by the (classical) AQUAL action.\footnote{A k-essence field is a scalar field whose Lagrangian depends non-linearly on the kinetic term, allowing for rich dynamical behaviour. It is typically used in modifications of gravity and to model the dynamics of dark energy and cosmic acceleration (Armendariz-Picon et al. [\citeyear{armendariz2001}]).} Similarly to the classical case, varying the action with respect to the scalar field yields a relativistic modified version of the Poisson equation:

\begin{equation}\label{relModPoisson}
    c^2\nabla.[f'(X)\nabla \phi] = kG\rho
\end{equation}
from which the relativistic interpolating function

\begin{equation}
    \tilde{\mu}(s)=(4\pi c^2/k)f'(X)
\end{equation}
is obtained, where $f(X)$ is the so-called `MOND function’. The relativistic interpolating function is related to the non-relativistic one through: $ \tilde{\mu}(s) = (x-s)s^{-1}$, with $ s=x[1-\mu(x)]$. In the deep-MOND regime, where $x \ll 1$, one gets $\tilde{\mu}(s) \sim s$ (since $s=x(1-x) \ll 1$ and $x\sim s(1+s)$), and therefore, even though $\tilde{\mu}(s)$ is generally different from its non-relativistic counterpart $\mu(x)$, it exhibits the same low-gravity asymptotic behaviour.

A later, and better-known attempt to construct a relativistic extension of MOND is Bekenstein's ([\citeyear{bekenstein2004}]) Tensor-Vector-Scalar theory (TeVeS). Compared to RAQUAL, TeVeS introduces a new vector and employs a disformal relation between the Einstein and physical metrics (as opposed to a conformal relation in RAQUAL) with the aim of addressing various difficulties in predicting gravitational lensing.\footnote{In short, the Einstein metric appears in the Einstein-Hilbert action after rewriting the theory in a GR-like form, while the physical metric couples directly to matter fields.} For our purposes, it suffices to note that TeVeS preserves RAQUAL's k-essence action for the scalar field, albeit with a different $f(X)_{TeVeS}$ function. The key idea remains the same: an appropriate function is freely chosen such that its derivative yields the required relativistic modified Poisson equation (\ref{relModPoisson}). The relation between this free function and Milgrom's non-relativistic interpolating $\mu-$function is the same as in RAQUAL (see also Famaey et al. [\citeyear{famaey2007}], Sec. III).

As is well-known, k-essence relativistic extensions of MOND such as RAQUAL and TeVeS face significant challenges, primarily due to the prediction of super-luminal propagation of gravitational waves and difficulties in accounting for gravitational lensing effects (Soussa and Woodard [\citeyear{soussa2004}]). MOND scholars have proposed various possible solutions; however, these issues need not concern us here. What matters for our analysis is how the discussed classical and relativistic versions of MOND, namely AQUAL, RAQUAL and TeVeS, reduce to Newtonian gravity through mathematical limits using an interpolating function, and whether these limiting reductions can indeed be justified according to the three criteria formulated above.

For completeness, let us also note that Bekenstein's original theory was later generalized by \cite{skordis2008} and \cite{skordis2021} whose theory (often dubbed Relativistic MOND (RMOND) or Aether Scalar Tensor (AeST) theory) is the current working candidate for a relativistic version of MOND. Like TeVeS, AeST contains a function $J(Y)$ in the action whose derivative determines how the added field contributes to the potential so that the total field obeys MOND in low accelerations and recovers Newtonian dynamics at higher accelerations. Nevertheless, AeST's dependence on this function is less direct compared to TeVeS: instead of freely choosing an interpolation function, one specifies a more fundamental function $F(Y,Q)$ introduced in the relativistic Lagrangian, from which the interpolating function $J$ is derived. In a sense, the interpolating behaviour is thus not imposed directly as in RAQUAL and TeVeS but emerges from the choice of $F$, which is embedded one level deeper, that is, in the relativistic Lagrangian. While our analysis in what follows is based on RAQUAL and TeVeS, we acknowledge that this special feature of RMOND/AeST introduces an additional level of complexity that warrants further study in its own right.

\subsection{The interpolating function}

To fully understand how AQUAL, RAQUAL and TeVeS reduce to Newtonian gravity by virtue of mathematical limits, it is useful to clarify the properties and the exact role of the interpolating $\mu$-function and its relativistic counterpart. As noted above, the loose nature of constraints on the interpolating function provides room for a variety of possible functions. One of the most widely used interpolating functions is the `simple' $\mu$-function, which yields considerably accurate results in the intermediate to weak gravity regime of galaxies, as opposed to the strong gravity regime of the solar system (Famaey and Binney [\citeyear{famaey2005}]). The `simple' $\mu$-function has the form:

\begin{equation}
    \mu(x) = \frac{x}{1+x}
\end{equation}
where $x=(g/a_0)$.\footnote{Other examples of interpolating functions are the $\alpha$-family (Angus et al. [\citeyear{angus2006}]) with $\tilde{\mu}_a(s) = \frac{s}{1-\alpha s}$ for $0 \leq \alpha \leq 1$, and the n-family with $\mu_n(x) = \frac{x}{(1+x^n)^{1/n}}$ where the $n=2$ case is widely used to analyse rotation curves and known as the `standard' $\mu$-function (Begeman et al. [\citeyear{begeman1991}], Sanders and McGaugh [\citeyear{sanders2002}]).}

In the case of AQUAL, taking the limit $a_0 \rightarrow 0$ for the high-acceleration (Newtonian) regime, where $g=|\nabla\Phi| \gg a_0$, gives $\mu\left(\frac{|\nabla\Phi|}{a_0} \right)\approx 1$, and thus the modified Poisson equation (\ref{ModPoisson}) reduces to the standard Poisson equation ($\nabla^2\Phi=4\pi G\rho$). Similarly, taking the low acceleration limit $a_0 \rightarrow \infty$ for the deep-MOND regime, where  $g=|\nabla\Phi| \ll a_0$, gives $ \mu\left(\frac{|\nabla\Phi|}{a_0} \right)\approx x$, and the modified Poisson equation (\ref{ModPoisson}) provides the desired MOND dynamics for which $g_M \sim \sqrt{g_Na_0}$. The interpolating function is therefore necessary to ensure a smooth and mathematically valid transition between the MOND and Newtonian regimes. 

For RAQUAL, things are slightly more complicated, however, the overall framework remains the same. The additional complication is that the relativistic theory must be reducible both to the intermediate Newtonian regime of low accelerations and the required MOND dynamics in the ultra-weak limit of very low accelerations below $a_0$. The reduction to MOND dynamics is achieved by taking the ultra-low acceleration limit $a_0 \rightarrow \infty$, where $X \ll 1$, and fine-tuning $f(X)$ so that $f'(X) \sim X^{1/2}$, to obtain the desired MOND law ($|\nabla\phi| \sim g_M \sim \sqrt{g_Na_0}$) from the relativistic modified Poisson equation (\ref{relModPoisson}). In this case, the relativistic interpolation function is responsible for tuning the scalar field dynamics to recover a MOND behaviour from RAQUAL in the low acceleration regime.

Similarly, the reduction to the Newtonian regime is achieved by taking the limit $a_0 \rightarrow 0$ and fine-tuning $f(X)$ such that $f'(X) \rightarrow \textit{constant}$. This makes the additional scalar field $\phi$ decouple and its contribution to the physical metric $g_{00} = -e^{2(\Phi+c^2\phi)/c^2}$ becomes negligible.\footnote{In general relativity, the component $g_{00}$ is the time-time component of the spacetime metric. It determines the gravitational time dilation experienced in a gravitational field, as well as the potential energy of particles. The scalar field $\phi$ introduced in RAQUAL modifies the gravitational potential via the physical metric, resulting in the above modified expression.} The physical metric thus becomes dominated by the standard Newtonian gravitational potential $\Phi$ reproducing the standard dynamics of General Relativity with $g_{00} = -e^{2\Phi/c^2}$, from which the Newtonian behaviour can be obtained by the standard limiting reduction (see Batterman [\citeyear{batterman1995theories}], Fletcher [\citeyear{fletcher2019reduction}]). 
This process ensures that for accelerations much larger than $a_0$ the predictions of RAQUAL converge with the predictions of general relativity, and consequently those of Newtonian gravity in the weak-field limit. In terms of the relativistic interpolating function, the particular choice of $f(X)$ for the Newtonian regime means that $\tilde{\mu}(s) \sim 1$ for $s \gg 1$.

TeVeS reduces to MOND and Newtonian gravity in a very similar manner, with only a few differences in the details due to the introduction of an additional vector field to the metric. In the ultra-low acceleration regime, the scalar field dominates the dynamics and the relativistic equations are once again fine-tuned via the choice of an appropriate function $f(X)_{TeVeS}$ in the scalar field action, ensuring the reproduction of MOND behaviour. In the high-acceleration (Newtonian limit) both the scalar and vector field contributions become negligible, resulting in the dynamics being governed only by the metric tensor. In this limit, the physical metric reduces to the standard metric of general relativity (just like in the case of RAQUAL) and the theory reproduces the weak-field Newtonian dynamics in the standard way.

In sum, MOND theories aim to reproduce Milgrom's law (\ref{Milgrom'sLaw}) by maintaining a smooth transition between  relativistic, newtonian, and deep-MOND regimes. In classical models like AQUAL, the Newtonian limit arises by taking $a_0 \rightarrow 0$ and using an interpolating function. In relativistic theories such as RAQUAL and TeVeS, the deep-MOND limit is obtained by taking $a_0 \rightarrow \infty$ and fine-tuning the interpolation function to obtain Milgrom's law, while the Newtonian regime follows indirectly in the standard way without interpolation.\footnote{This section mainly draws on \cite{famaey2012}. For a more recent review of MOND theory with a focus on observational evidence, see also \cite{banik2022}.}

\section{MOND as a valid (?) case of limiting reduction}\label{MONDLRSEC}

Having shown how MOND theories use an interpolating function to smoothly transition between regimes via mathematical limits, we now assess whether these cases meet the three criteria for a successful limiting reduction: reduction via a mathematical limit, plus the additional conditions of physicality and the non-arbitrariness of the physical parameter.

Recall that according to \cite{nickles1973} and \cite{palacios2022} a quantity $Q^2$ of a theory $T^2$ reduces to a corresponding quantity $Q^1$ of a theory $T^1$ iff taking the limits of $Q^2$ with respect to a given parameter $N$ yields $Q^2$, insofar as the limiting operation makes physical sense. Leaving the second clause on the side, we can clearly see that the three aforementioned cases of limiting reduction from AQUAL to Newtonian gravity, from RAQUAL (and TeVeS) to AQUAL and from RAQUAL to general relativity (and subsequently to Newtonian gravity) clearly fit Nickels' account of limiting reduction. In particular, in all three cases, the quantity of true acceleration (MOND-ian), $g_M$, reduces to Newtonian acceleration, $g_N$ taking the lower limit of the acceleration constant $a_0$, or equivalently, taking the upper limit of the parameter $x$ in the interpolating function. In Palacios' ([\citeyear{palacios2022}]) formulation, this can be written as:

\begin{equation}\label{MONDLR}
\lim_{a_0\to 0} g_M = g_N \ \ \ \ \text{or} \ \ \ \ \lim_{x\to \infty} g_M = g_N
\end{equation}

To fully qualify as a successful case of limiting reduction, one needs to further ask whether the employed limit `makes physical sense'---that is, whether the mathematical limit satisfies the additional two conditions of \textit{physicality} and \textit{non-arbitrariness}. We analyse them in turn in the context of MOND and argue that the former is not problematic (although the issue is more complex than it may seem) whereas the latter aspect is more troubling and threatens the justification of MOND theories. The latter aspect is elaborated on further in Section \ref{Planck}.

Let us begin with physicality. As we have seen, a possible way to address the issue of physicality is by applying Butterfield's principle according to which a limit is justified if the values of the quantities in the limit approximate the values of the quantities `on the way' to the limit, that is, if $Q_{\infty} \approx Q_{N_0}$ (for large but finite values $N_0$ of the limiting parameter), and the behaviour of the theory on the way to the limit is physically realistic. In the case of limiting reduction in MOND theories, this amounts to the requirement that as the theory goes to higher accelerations, where $g \gg a_0$, `true' acceleration smoothly becomes equal to Newtonian acceleration and vice versa; as the theory goes to lower accelerations, where $g \ll a_0$, `true' acceleration shifts away from Newtonian acceleration and gradually changes according to Milgrom's law. While (\ref{MONDaccel}) clearly fails to satisfy this requirement, the interpolating function in Milgrom's law (\ref{Milgrom'sLaw}) is introduced precisely for this purpose. That is, to ensure that the gravitational acceleration smoothly approximates the value of Newtonian acceleration on the way to the $a_0 \rightarrow 0$ / $x \rightarrow \infty$ limit, and the value of the MONDian acceleration changes on the way to the $a_0 \rightarrow \infty$ / $x \rightarrow 0$ limit in a physically realistic way, that is, without any fluctuations or abrupt changes.

Moreover, the distinct nature of limiting reduction in MOND theories is revealed once we ask the further question of whether the involved limits are regular or singular. Recall that for some authors (Batterman [\citeyear{batterman2001}], Rueger [\citeyear{rueger2000, rueger2004}]) the presence of singular limits indicates a failure of limiting reduction between two theories. The possible presence of a singular limit in the reduction of MOND theories to Newtonian gravity would therefore indicate a possible failure since in such cases Butterfield's principle is not satisfied. Within the MOND framework, this would amount to a case in which the value of gravitational acceleration as $x \rightarrow \infty$ does not converge to the quantity evaluated at the limit $x = \infty$, that is, in cases where $\lim_{x\to \infty} g_N \neq g_{x=\infty}$.

\textit{Prima facie}, the global limit $a_0 \rightarrow 0$ is singular precisely because the fundamental dependence of MOND theories on the acceleration constant, $a_0$, means that MONDian gravitational behaviour disappears completely when $a_0 \rightarrow 0$, implying an abrupt change in the behaviour of the theory. In simple words, when $a_0 \rightarrow 0$ the MONDian behaviour suddenly disappears and gravitational dynamics are described by the different laws of Newtonian gravity. In this regime, the interpolation function no longer serves as a transition bridge between the Newtonian and deep-MOND regimes, since the theory completely reduces to Newtonian dynamics as $\mu(x)$ becomes approximately equal to 1. This fundamental change in the behaviour of the theory when the acceleration constant vanishes might be taken as an indication that the limit is singular, especially if one ignores, for a moment, the crucial role of the interpolating function. Moreover, the potential singularity of the $a_0 \rightarrow 0$ limit can be shown by realising that gravitational acceleration cannot be expanded as a Taylor series unless the form of the interpolation function is clearly specified. Recall that in the ultra-low acceleration (MONDian) where one takes the $a_0 \rightarrow \infty$ limit and $\mu(x) = x$, the MOND equation becomes:

\begin{equation}
    g\frac{g}{a_0} = g_N \implies \ \ g=\sqrt{g_Na_0}
\end{equation}
indicating a non-analytic behaviour in $a_0$ since the acceleration $g$ depends on $\sqrt{a_0}$. This means that, unlike the example of momentum in special relativity, the reduced quantity here (the gravitational acceleration, $g$) cannot be expressed as a smooth Taylor expansion in $a_0$ since it involves a fractional power of $a_0$ which is not analytic and  therefore, the limit is singular.

What is so special about MOND however, is that the determination of a specific form for the interpolating function with certain properties renders the $a_0 \rightarrow 0$ limit regular. To see why, consider the `simple' $\mu$-function for which $\mu(x)=\frac{x}{1+x}$ and $x=g/a_0$. Plugging this function in Milgrom's law, (\ref{Milgrom'sLaw}) and carrying out some simple algebraic simplifications yields the following relationship between MONDian, $g_M$ and Newtonian, $g_N$, acceleration:

\begin{equation}\label{SimpleFunction}
    \frac{g_M^2}{a_0+g_M} = g_N
\end{equation}

When $a_0 \rightarrow 0$, the equation becomes $g_M=g_N$, which means that as $a_0 \rightarrow 0$, $g_M$ smoothly approaches $g_N$ without any discontinuities or divergences and hence the limit becomes regular. What is more, (\ref{SimpleFunction}) can be expanded into a Taylor series around $a_0 = 0$ which can be taken as a justification that the limit $a_0 \rightarrow 0$ is actually regular. In particular, solving for positive $g$ yields:

\begin{equation}
    g_M = \frac{g_N + \sqrt{g^2_N +4g_Na_0}}{2}
\end{equation}
which can be expanded as a Taylor series of $g$ around $a_0=0$ as:

\begin{equation}
    g = g_N +a_0-\frac{a^2_0}{2g_N}+\mathcal{O}(a^3_0)
\end{equation}
ensuring higher-order terms ($a_0,a^2_0,...$) vanish smoothly as one approaches Newtonian regime. 

While the generic MOND formulation and the abrupt loss of the deep-MOND regime as $a_0 \to 0$ suggest a singular limit, introducing a specific interpolating function regularizes it, ensuring a smooth, mathematically justified transition between regimes. Thus, the limiting reduction of MOND to Newtonian gravity also satisfies \textit{physicality}.

The condition of \textit{non-arbitrariness} is the most crucial. Recall that non-arbitrariness, as formulated above based on Nickles' account, requires that no arbitrary factor, to which we cannot give a physical interpretation and on which the limit is taken, is introduced to recover the narrower theory. The crucial question to ask here is whether the limiting reduction in MOND involves some kind of arbitrariness in the above sense. 

A natural and general point could be that the \textit{ad hoc} introduction of an interpolating function by hand is, in principle, profoundly arbitrary. We shall leave this objection aside for a moment and argue that there is a clear sense in which the limiting reduction in MOND is non-arbitrary. We shall then return to the \textit{ad-hoc} nature of interpolation and show that, insofar as it can be given a physical meaning, an arbitrary act of interpolation between different regimes is not pathological by itself. To support our argument, we will employ a well-known historical example of interpolation between different regimes in the next section, namely the derivation of Planck's law from Wien's and Rayleigh's laws. Before that, let us first see how the view that limiting reduction in MOND is not arbitrary can be supported.

To begin with, note that there is a clear sense in which the limiting factor itself, $a_0$ (or equivalently the parameter $x=g/a_0$), has a clear physical interpretation as a fundamental physical constant with a well-defined value that arises naturally from various scaling relations in galactic dynamics. As often argued by MOND scholars, the acceleration constant can be understood as playing a similar role to the Planck constant, $h$, in quantum mechanics and the speed of light $c$ in special relativity, in delimiting the different scales in which a theory can be applied. Milgrom's own words on this are rather illuminating: 

\begin{quote}
    It can be said then that $\hbar$, $c$, and $a_0$ all play a role of ‘boundary constants’, or delimiters of the applicability regime of the old paradigm, in whose equations they do not appear. Classical Newtonian dynamics is the common limiting paradigm of all three when $\hbar \rightarrow 0$, $c \rightarrow \infty$, $a_0 \rightarrow 0$. (Milgrom [\citeyear{milgrom2020mond}], p. 172)
\end{quote}

Hence, whether or not MOND is correct, the limiting factor in recovering Newtonian dynamics is well-defined and physically meaningful, technically satisfying non-arbitrariness.

Nevertheless, it can be said---and this is one of the most common objections against MOND---that the loose nature of the constraints on the interpolating function implies a kind of arbitrariness, as one is essentially free to introduce whatever form of the interpolating function satisfies the three basic constraints. As already mentioned, there are currently several viable candidates for the exact form of the interpolating function, and in many cases different functions seem to fit different kinds of data, giving the sense of an arbitrary fine-tuning in the choice of the exact form. However, while this is true, the underdetermination of the exact form of the interpolating function can, in principle, be attributed to the lack of precise data that could potentially pin down its exact form. In other words, the arbitrariness in the exact form of the interpolating function can be seen as an epistemological limitation due to the lack of data, and the current uncertainty will essentially vanish once its exact form is determined by more precise observations. In this sense, the final form of the interpolation is not arbitrary.

However, more generally, one might argue that the requirement to introduce no arbitrary factor to recover the narrower theory should extend not only to the new limiting parameter but also to the interpolating function. The underlying idea is that the transition between the two theories should not be achieved in an arbitrary way by tailoring the form of the newly introduced theory to connect in the right way to the existing one. Since the interpolating function in MOND theories is introduced exclusively to recover Newtonian dynamics, it is inherently arbitrary and renders the reduction to Newtonian gravity unjustifiable. This is a plausible objection, and we agree that the interpolating function is introduced in MOND theories with the exclusive role of bridging between different regimes.\footnote{Our analysis here is in line with a recent and broader discussion on MOND's overall ad-hocness/arbitrariness by \cite{duerr2023}. There, the authors conclude that the introduction of a free function, which is tantamount to the introduction of infinitely many free parameters, diminishes the theory's coherence with theoretical background knowledge (p. 13).} However, this argument does not necessarily lead to the conclusion that the limiting reduction is pathological. In the next section, we shall use the well-known example of Planck's law of electromagnetic radiation to show that, insofar as the limiting factors make physical sense, an ostensibly arbitrary interpolation between different regimes can, in principle, be justified by appealing to a more fundamental theory. Hence, although the introduction of an interpolating function in MOND theories is indeed arbitrary, the potential pathology of the limiting reduction to Newtonian dynamics should not be attributed to this fact alone, rather it should be sought in the absence of a more fundamental theory and the fact that there is no common and unified structure between the different regimes in which such theory potentially applies (more in Section \ref{GR_Vs_MOND}).

\section{Limiting reduction and Arbitrariness: Planck's example}\label{Planck}

A useful historical example with notable similarities to the interpolating nature of MOND is the derivation of Planck's law of black-body radiation in the early 1900s. Similarly to what happens in MOND, Planck's law was devised in a rather \textit{ad hoc} manner by arbitrarily interpolating between different regimes based on the existing phenomenological---albeit flawed---laws proposed by Wilhelm Wien and Lord Rayleigh. What makes the two situations different? As we shall see, although Planck derived his law by interpolation with the sole aim of reconciling the two laws, the final result is a valid example of limiting reduction between regimes, because Planck's law was later independently derived on solid theoretical grounds and was eventually justified by the theory of quantum mechanics.\footnote{In presenting Planck's law and its origins we largely follow \cite{duncan2019constructing}.} To distinguish the two situations, we suggest that Planck's case---as opposed to MOND's case---satisfies an additional non-arbitrariness criterion for successful limiting reductions: `\textit{downward theoretical grounding}'.

The starting point for black-body radiation physics was Wien’s law, which was derived in 1893 based on entropy considerations in thermodynamics. Wien's law related the energy, $\rho_\nu$, per unit volume with frequency, $\nu$, and temperature, $T$:

\begin{equation}\label{Wienslaw}
  \rho_\nu = c_1\nu^3e^{-c_2\nu/T}
\end{equation}
where $c_1$ and $c_2$ are free parameters to be obtained experimentally. The law initially provided an excellent approximation for blackbody radiation; however, the gradual accumulation of experimental results eventually began to challenge its validity. For instance, soon after the initial formulation of the law by Wien, Lummer and Pringsheim ([\citeyear{lummer1900strahlung}], p. 171) pointed out the incompleteness of the law in the mid-infrared spectrum based on their experimental results from blackbody radiation for long waves: ``the Wien–Planck spectral equation does not represent the black radiation measured by us in the range from \(12\mu\) to \(18\mu\).", where \(\mu\) is the wavelength.\footnote{As quoted in Kangro ([\citeyear{kangro1976}], pp. 195-6) and Duncan and Janssen ([\citeyear{duncan2019constructing}], p. 67).}

Meanwhile, an alternative to Wien’s law was also proposed in June 1900 by Lord Rayleigh, based on entirely different principles. Employing arguments from classical physics and the equipartition theorem in statistical mechanics, he found that the energy density, \(\rho_\nu\), of black-body radiation is proportional to \(\nu^2T\), where \(\nu\) is the frequency and \(T\) is the temperature:

\begin{equation}\label{rayleigh's law}
    \rho_\nu = \frac{8\pi}{c^3}\nu^2kT
\end{equation} 
However, Rayleigh soon recognized that this law could not hold for all values of \(\nu\): as frequency increased from the infrared to the visible and ultraviolet, energy would grow without bound, leading to what is often called as the `ultraviolet catastrophe'.  

To obtain a viable law for the entire spectrum, Rayleigh, somewhat arbitrarily from a theoretical perspective, multiplied the expression \(c_1\nu^2T\) by a function \(e^{-c_2\nu/T}\), where \(c_1\) and \(c_2\) are constants to fit the experimental data. However, experimental results indicated that while Rayleigh’s law significantly improved Wien’s law at longer wavelengths, it performed much worse at shorter wavelengths, leading to an uncomfortable situation in which two seemingly unrelated and incomplete laws accounted for blackbody radiation in different regimes.

To reconcile the two laws, Planck first revised Wien’s law by proposing a new equation for the second-order derivative of the entropy, $S$, of a resonator as a function of its energy, $U$:  

\begin{equation}\label{planck1}
    \frac{\partial^2 S_\nu}{\partial U^2_\nu} = -\frac{\alpha}{U_\nu(\beta+U_\nu)}
\end{equation}
Just like in MOND, the sole aim of this equation was to smoothly interpolate between the two regimes of black-body radiation described by the two laws by Wien and Rayleigh and achieve the best of two worlds, albeit in a rather \textit{ad hoc} and arbitrary manner. Eq. (\ref{planck1}) then provided the basis for the derivation of Planck’s law for black-body radiation in its current familiar form:

\begin{equation}\label{planck2}
    \rho_\nu = \frac{8\pi}{c^3}h \nu^3 \frac{1}{e^{h\nu/kT} - 1}
\end{equation}
in which Planck's famous constant $h$ was introduced ($k$ stands for the Boltzmann constant).

Given that Planck’s law was explicitly constructed to interpolate between the two existing laws, it is no surprise (\ref{planck2}) reduces to Wien’s law in the high-\(\nu\) (low-\(T\)) limit, where \(e^{h\nu/kT} \gg 1\):  

\begin{equation}
    \rho_\nu \approx \frac{8\pi}{c^3} h\nu^3 e^{-h\nu/kT}
\end{equation} 
and to Rayleigh's law in the low-\(\nu\) (high-\(T\)) limit, where \(e^{h\nu/kT} \approx 1 + h\nu/kT\):  

\begin{equation}\label{ray}
    \rho_\nu \approx \frac{8\pi h\nu^3}{c^3}  \frac{kT}{h\nu} = \frac{8\pi}{c^3}\nu^2kT
\end{equation}  

Let us take the latter relation as an example and reformulate it to highlight how this instance of reduction fits within the account of limiting reduction we presented. In the limit where $h\nu/kT \rightarrow 0$ the exponential function in Planck's law (\ref{planck2}) can be expanded to a Taylor series from which we can keep only the first two terms $1 + h\nu/kT$.\footnote{For very small values of \( x = \frac{h\nu}{kT} \), we can expand the exponential in a Taylor series:
\(e^x \approx 1 + x.\)} Inserting these terms into (\ref{planck2}) we obtain Rayleigh's law (\ref{rayleigh's law}) (Bokulich [\citeyear{Bokulich2008-BOKRTQ}], §1.4). This procedure ensures that both conditions (a.)--(b.) for a successful limiting reduction are satisfied, that is, \textit{limit} and \textit{physicality}. Condition (a.) (\textit{limit}) is met because the (classical) energy density in Rayleigh's law is recovered from the quantum one in the limit. Condition (b.) (\textit{physicality}) is satisfied because the limit is reached in a non-singular way via the expansion operation, ensuring that Butterfield's principle is satisfied and the limit makes physical sense.

The condition of \textit{non-arbitrariness} is again the most crucial. First, note that the limiting parameter, $h$, was given a clear physical interpretation through Planck's famous suggestion that energy is quantized and could only be emitted or absorbed in discrete amounts; Planck's constant related the energy $E$ of a quantum of electromagnetic radiation to its frequency $\nu$ through the formula $E=h\nu$. The derivation of his law for black-body radiation was essentially a combination of his idea of quantized energy and the principles of statistical mechanics, hence the limiting parameter $h$ had a solid and meaningful physical interpretation.

Nevertheless, what makes Planck's example particularly relevant to our discussion of limiting reduction in MOND is the fact that Planck initially derived his law with the sole aim of interpolating between Wien's and Rayleigh's laws, without any solid theoretical justification. Hence the crucial question is whether the \textit{ad hocness} of Planck's derivation is by itself sufficient to question the status of the law as a successful case of limiting reduction. Planck's own concerns in his autobiography years later for the theoretical justification of his law despite its empirical success are particularly illuminating:

\begin{quote}
    ...so long as it had merely the standing of a law disclosed by a lucky intuition, it could not be expected to possess more than a formal significance. For this  reason [...] I began to devote myself to the task of investing it with a true physical meaning (Planck [\citeyear{Planck1949-PLASAA-3}], p. 41).
\end{quote}

Planck's endeavour to find the true physical meaning of the law would indeed soon come to an end through the justification of his law in the broader context of quantum theory. The principle of the quantisation of energy which was initially introduced as a useful mathematical tool eventually became the cornerstone of quantum mechanics and was fully justified in subsequent theoretical developments. Crucial to these developments were Einstein's seminal explanation of the photoelectric effect based on the quantisation of energy and the independent derivation of Planck's law in 1917  from his quantum theory of radiation \citep{einstein1917quantentheorie}. Starting from non-classical assumptions about the quantisation of energy of molecules and the exchange of energy and momentum between molecules and light quanta, and assuming Wien's law, Einstein was able to derive the Planck radiation law (up to a constant) (Duncan and Janssen [\citeyear{duncan2019constructing}], §3.6). Moreover, Planck's law was also derived in 1924 from the Bose-Einstein distribution---a statistical framework for bosons---providing even more justification to what initially looked like an intuitive mathematical exercise of interpolation. By calculating the energy density of a photon gas in thermal equilibrium, Bose and Einstein managed to reproduce Planck's law without involving any kind of interpolation. Their derivation essentially solidified the theoretical foundation of blackbody radiation in quantum mechanics, removing any possible doubts on the status of Planck's law as a successful case of limiting reduction.

The upshot is that while the initial derivation of Planck's law was indeed an arbitrary act of interpolation in the sense just presented, the law was later fully justified and it is now seen as an uncontroversial case of a valid limiting case between Wien's and Rayleigh's laws. Thus, insofar as the limiting parameters have a clear physical interpretation, arbitrarily introducing such parameters to achieve the reduction of a theory to a narrower theory does not necessarily render the limiting reduction pathological. 

The crucial difference between Planck's example and MOND is the eventual justification of the \textit{ad hoc} interpolation by a more fundamental and solid theoretical framework. Planck's law was later explained and validated by the principle that energy exchange between matter and electromagnetic radiation is quantized, and received further validation by being consistent with the other laws of quantum mechanics. Hence, the law obtained by interpolation is justified because we can explain it via a more fundamental or more broadly scoped theory. 

On the contrary, the introduction of the interpolating function in classical and relativistic theories of MOND has not been explained and validated in the same way within a fundamental theory. Current versions of MOND essentially lack this kind of justification and this is precisely one of the main reasons that renders the limiting reduction in this case pathological. The need for a deeper justification of the interpolation function via a more fundamental theory (usually termed FUNDAMOND) is also often expressed by the father of MOND, Mordehai Milgrom, who, like Planck, openly recognizes the shortcomings of the unjustified introduction of the interpolating function to achieve the transition to different regimes:

\begin{quote}
All MOND theories proposed to date introduce by hand the interpolation between standard dynamics and the DML [deep-MOND limit] as the acceleration goes from high to low. Clearly, we need an underlying theory in which such interpolation will emerge from the theory itself. (Milgrom [\citeyear{milgrom2020mond}], p. 177)
\end{quote}

Some MOND scholars go even further by claiming that not only the interpolating function will be justified via a more fundamental theory of gravity, but it will also vanish:

\begin{quote}
		The very concept of a pre-defined interpolating function should even in principle fully disappear once a more profound parent theory of MOND is discovered. (Famaey and McGaugh [\citeyear{famaey2012}], p. 53)
\end{quote}

Hence the pathology of the limiting reduction in the case of MOND is not to be found in the bare fact that the interpolation function is introduced to achieve the transition to the Newtonian regime. Planck used a very similar strategy, but the result of his strategy is nonetheless an uncontroversial example of a successful limiting reduction. Rather, what makes the MOND cases problematic is the absence of a fundamental theoretical framework to justify the interpolation. To put it more briefly, MOND fails to satisfy a specific condition of non-arbitrariness not originally specified by the original criteria (a.)--(c.), which we call `\textit{downward theoretical grounding}'. Section 7 elaborates on the relationship of this condition with the original criteria, highlighting how this can be taken as a refinement of condition (c.).

It is worth noting that the kind of dialectical reasoning through which Planck's law was justified (via downward theoretical grounding) occurs widely in the history of physics and is not restricted to cases involving interpolation, as in Planck's law or MOND. Notable examples can be found in semiclassical mechanics.\footnote{For an introduction to semiclassical mechanics, especially within the context of theoretical reduction, see Bokulich ([\citeyear{Bokulich2008-BOKRTQ}], Ch. 5). For a physics reference, see \cite{berry1972semiclassical}.} Within this field it is often the case that the starting point is classical laws (like the Hamilton--Jacobi equation), which are heuristically modified to account for quantum phenomena (somewhat analogously to how classical laws are modified in MOND), and which in turn find their justification in the full framework of quantum theory. To go into more detail, in the early 20th century physicists struggled to understand why atoms exhibited discrete spectral lines rather than the continuous spectra predicted by classical mechanics. Classical orbits and Maxwell’s electrodynamics simply could not explain atomic stability or the observed regularities in emission and absorption frequencies. Faced with this gap, theorists sought ways to `quantise' classical motion---not by abandoning classical mechanics entirely, but by modifying it with new rules that imposed discreteness where none had existed. This effort gave rise to what came to be known as the \emph{old quantum theory}.

The Bohr--Sommerfeld rules were its most influential tool. They began as a pragmatic repair: electrons were still assumed to follow definite classical orbits, but only those orbits whose action integrals satisfied
\(
\oint p\,dq = n h
\)
were allowed. This \textit{ad hoc} condition introduced quantized energy levels and, in Sommerfeld’s hands, even explained fine details such as the hydrogen atom’s fine structure. Yet the rule itself was heuristic. It was not derived from deeper principles but rather justified by empirical adequacy: it `worked' for hydrogen and a few other simple systems, even as it failed badly for more complex atoms. In this sense, the old quantization rules were not a coherent new mechanics, but a patchwork that fused classical dynamics with quantization conditions chosen largely for their empirical success.

Over the next decades, however, these rules were improved and placed on firmer ground. The development of wave mechanics made clear that the half-integer shifts missing from Bohr--Sommerfeld quantization arose from the wave nature of particles: near a classical turning point, a wavefunction must match smoothly between oscillatory and decaying regions, and this inevitably introduces a universal phase shift of $\pi/2$. The WKB method captured this correction, and Einstein, Brillouin, and Keller generalised it into what is now called EBK quantization, in which additional terms---the Maslov indices---systematically account for all such phase shifts and boundary effects. EBK conditions not only repaired the deficiencies of the old rules but also reproduced exactly the spectra of benchmark systems like the harmonic oscillator and hydrogen atom. What began as a heuristic compromise with classical mechanics was thus revealed, in retrospect, to be the semiclassical limit of a broader and more rigorous quantum framework.

\section{Limiting reduction and Arbitrariness: General Relativity}\label{GR_Vs_MOND}

In Section \ref{Planck} we compared the application of limiting reduction to Planck's law with the case of MOND, drawing important considerations about the reduction-wise justification of MOND and limiting reduction in general. In this section, we shall compare MOND to the reduction of general relativity to Newtonian gravity and draw further conclusions along the same lines. The case of general relativity is particulalry helpful because it (i) further clarifies the specific sense in which MOND can be regarded as failing to be reduction-wise justified, and (ii) it points to an additional refinement of the original non-arbitrariness condition—beyond what is illustrated by Planck’s case—namely the requirement of \textit{upward theoretical grounding}, which helps ensure that instances of limiting reduction are genuinely justified.

Let us recap. Section \ref{MONDLRSEC} argued that a widely considered pathological case of limiting reduction, that is, the liming reduction of classical and relativistic versions of MOND to Newtonian gravity, formally satisfies the standard conditions for liming reduction set by Nickles' original account in the following sense: (a.) $g_M$ reduces to $g_N$ by taking an appropriate limit of a parameter ($x=g/a_0$) (\textit{limit}); (b.) the limiting operation involves a regular limit and the quantity can be expanded in a Taylor series (Butterfield's criterion $\rightarrow$ \textit{physicality}); (c.) the limiting parameter is not arbitrarily introduced and can be given a physical interpretation (\textit{non-arbitrariness}).

Regarding non-arbitrariness, we argued that while the limiting parameter $a_0$ is not arbitrarily introduced and has a clear physical meaning, the interpolating function enabling regime transitions is introduced arbitrarily. Yet, as the Planck-law case shows (§\ref{Planck}), interpolation alone need not make a limiting reduction pathological; what matters is whether it can be grounded in a more fundamental theory. This distinction separates MOND from Planck’s case, showing that conditions (a.)--(c.) are too coarse to judge MOND’s status; the non-arbitrariness criterion hence requires further refinement.

To complete our refinement of limiting reduction, we finally consider yet another standard successful example of reduction---general relativity to Newtonian gravity---and compare it to MOND. We argue that MOND fails to satisfy a further condition that should be posed on limiting reductions to avoid arbitrariness, which is instead satisfied by general relativity.

General relativity is often taken as a prime example of a successful limiting reduction because the reduction to Newtonian gravity occurs systematically and naturally within a unified mathematical framework. The process of arriving at Newtonian dynamics by taking specific limits such as low velocities, weak gravitational fields, etc. arises from the internal structure of the theory and does not require any \textit{ad hoc} modifications or any special mathematical treatment to retrieve it. Most importantly, the mathematical framework of Newtonian gravity is already embedded in the internal structure of general relativity and thus arises naturally when certain limits are taken. One can therefore imagine a possible scenario in which general relativity is introduced first, and from which one can derive an effective, and considerably simpler theory of gravity (Newtonian gravity) for low velocities. In other words, the mathematical structure of general relativity as the broader (reduced) theory does not depend on the mathematical structure of the narrower (reducing) theory of Newtonian gravity.

Modified Newtonian Dynamics and its relativistic extensions do not stand on the same ground. They are, for the time being, phenomenological theories crafted to reproduce Newtonian dynamics at a certain scale, and their mathematical framework is thus inherently dependent on the structure of Newtonian gravity. In other words, as opposed to general relativity which offers a unified mathematical framework for all scales into which the framework of Newtonian gravity is naturally embedded, MOND theories in their current form are essentially an artificial `patch' of two distinct mathematical frameworks, which is precisely what makes the introduction of the troublesome interpolation function necessary. The key difference between general relativity and MOND is therefore the fact that the former provides a unified mathematical structure for different scales into which the narrower theory is embedded. This means that, although Newtonian gravity provides a much simpler framework to solve problems in the weak field limit, the overarching mathematical framework of general relativity can still be applied without necessarily resorting to the narrower theory. This is not the case in MOND however. Although a classical theory of MOND is a broader theory than Newtonian gravity, when the $a_0 \rightarrow 0$ limit is taken one still needs to resort to the mathematical framework of the Newtonian gravity. As opposed to general relativity, it is therefore harder (methodologically speaking) to imagine a scenario in which a classical or relativistic version of MOND preexists the narrower theory of Newtonian gravity and to which the mathematical framework of the latter is naturally embedded. This is another aspect that is not captured by the standard understanding of limiting reduction. We suggest that this be understood as a further implementation of the non-arbitrariness requirement, which we call \textit{upward theoretical grounding}, as it will become clearer in the next section.

An important point to stress is that this requirement does not impose the stricter condition that the broader theory must avoid using the theoretical resources of the narrower theory in any way. Once again, semiclassical mechanics and its relation to quantum mechanics provides a good illustration. At the semiclassical scale, classical mechanics equations and concepts enter into the description of quantum systems in a meaningful way. However, unlike MOND, quantum mechanics is a broader framework whose general mathematical structure is not merely a patchwork of classical mechanics equations---analogously to the general relativity example.

\section{Conclusion: A refined account of limiting reduction}\label{conclusions}

Summing up, we close by proposing a refined version of the limiting account of reduction—which consequently provides an improved account of reduction-wise justification—by refining the original condition (c.). A physical theory is \textit{reduction-wise justified} in terms of mathematical limits if, when it is introduced in a domain where a more narrowly scoped established theory already exists, it satisfies the following conditions:

\begin{enumerate}
    \item[(a.)] \textbf{Limit}: its quantities reduce to corresponding quantities of a narrower theory by taking an appropriate upper or lower limit of a parameter.
    \item[(b.)] \textbf{Physicality}: limiting operations are physically realistic and involve only regular limits.
    \item[(c.)] \textbf{Non-arbitrariness}: the narrower theory is recovered non-arbitrarily, that is:
    \begin{itemize}
        \item \textbf{Non-arbitrary limiting parameter:} if a new limiting parameter is introduced, it is not arbitrarily introduced and can be given a meaningful physical interpretation.
    \item \textbf{Downward theoretical grounding}: the transition between the two regimes is obtained in a non-arbitrary way, meaning that the form of the laws of the newer broader theory and any interpolating function introduced to recover the narrower theory should be justified by more general independent theoretical grounds, such as derivation from a more fundamental theory.
    \item \textbf{Upward theoretical grounding}: the transition between the two regimes is obtained in a non-arbitrary way, meaning that the framework of the narrower theory must be naturally embedded in the framework of the broader theory being introduced, and the latter should be able to account, in principle, for phenomena at all relevant scales without resorting to the framework of the narrower theory.
    \end{itemize}
\end{enumerate} 

Condition (c.) was inspired by Nickles’ original proposal, with the first point—\textit{non-arbitrary limiting parameter}—being explicitly invoked in the original account. This non-arbitrariness requirement provided an ideal starting point for our discussion of the validity of MOND \textit{vis-à-vis} limiting reduction. However, as we have seen, there are several senses in which an instance of reduction can be regarded as arbitrary, meaning that the original proposal was not sufficiently fine-grained. To address this, we refined (c.) by implementing two additional conditions within the non-arbitrariness criterion to ensure that potentially pathological or arbitrary instances of reduction are ruled out while allowing non-pathological cases, such as Planck’s law. The rationale for these two conditions has been heuristically motivated by the analysis of general relativity and the discussions in the preceding sections.\footnote{We emphasise that conditions (b.) and (c.) are related but distinct: (b.) requires a \textit{realistic} limit, ensuring smooth, non-singular transitions (see §2.2), but does not address the arbitrariness of how the limit is introduced, which must be controlled separately.}

Let's finally clarify what kind of conditions (a.)--(c.) are. We take condition (a.) as the necessary and defining feature of limiting reduction. On the other hand, we remain flexible concerning the status of (b.)--(c.). One option is to regard them as necessary conditions. In that case, MOND is simply reduction-wise unjustified, failing part of (c.). Another option, which we find less overly restrictive, is to take (b.)--(c.) as additional criteria that strengthen the justification of a case of limiting reduction. More precisely, criterion (b.) and the three conditions comprising (c.) should not be taken as necessary to achieve reduction-wise justification \textit{tout court}. Rather, it is possible to see justification as a matter of degree, where a theory can be more or less justified along different directions. Hence, conditions (a.)--(c.) above set an ideal standard for the fullest possible reduction-wise justification, but it might be that a theory is accepted even if the involved limits are not regular, or if there is no more fundamental theory or principle available to justify the form of its laws. For instance, one can conceive of the conditions in (c.) as setting a standard: the higher the standard of non-arbitrariness that is achieved, the more reduction-wise justified the theory is in that respect, and the more costly it would be to reject it. In this respect, we would conclude that MOND is not fully justified.

In conclusion, the MOND framework provides a useful case study for testing limiting reduction. The result is an improved version of the limiting account of inter-theoretic reduction, with possible implications well beyond the case of MOND.

\newpage

\section*{Acknowledgements}

Our heartfelt gratitude goes to Constantinos Skordis, as well as the audiences at the `Epistemology of Dark Matter and Modified Gravity' workshop in Athens and the Black Hole Initiative `Foundations seminars', for their valuable feedback. We are also grateful to the two anonymous reviewers at BJPS and to the editors for their helpful comments and suggestions. Antonis Antoniou is funded by the European Union’s Horizon 2020 Research and Innovation programme under the ERA Fellowship Action project ``ANDROMEDA'', Grant Agreement No. 101130750. Lorenzo Lorenzetti was supported by the Swiss National Science Foundation, SNSF Starting Grant project “Temporal Existence”, project number TMSGI1-211294.

\begin{flushright}
Antonis Antoniou \\ \textit{Department of History and Philosophy of Science} \\ \textit{National and Kapodistrian University of Athens} \\ \textit{Athens, Greece} \\ antantoniou@phs.uoa.gr

\vspace{0.5cm}

Lorenzo Lorenzetti \\ \textit{Institute of Philosophy} \\ \textit{USI -- Università della Svizzera italiana} \\ \textit{Lugano, Switzerland} \\ lorenzo.lorenzetti.ac@gmail.com
\end{flushright} 

\bibliography{library}

\end{document}